\documentclass{ws-procs975x65}

\def\AJ{{\it Astroph. J.} }
\def\AJL{{\it Ap. J. Lett.} }

\def\GRG{{\it Gen. Relativity and Gravitation} }

\def\JHEP{{\it JHEP} }

\def\MPL{{\it Mod. Phys. Lett.} }

\def\NAT{{\it Nature} }

\def\NP{{\it Nucl. Phys.} }
\def\PL{{\it Phys. Lett.} }
\def\PR{{\it Phys. Rev.} }
\def\PRL{{\it Phys. Rev. Lett.} }

\def\mev{\mbox{MeV}}

\def\frac#1#2{{\textstyle{{#1}\over {#2}}}}

\def\lsim{\mathrel{\rlap{\lower4pt\hbox{\hskip1pt$\sim$}}
    \raise1pt\hbox{$<$}}}
\def\gsim{\mathrel{\rlap{\lower4pt\hbox{\hskip1pt$\sim$}}
    \raise1pt\hbox{$>$}}}
\def\sqr#1#2{{\vcenter{\vbox{\hrule height.#2pt
         \hbox{\vrule width.#2pt height#1pt \kern#1pt
         \vrule width.#2pt}
         \hrule height.#2pt}}}}

\def\beq{\begin{equation}}
\def\eeq{\end{equation}}
\def\beqa{\begin{eqnarray}} 
\def\eeqa{\end{eqnarray}}

\begin{document}

\title{WMAP Constraints on a Quintessence Model \footnote{Talk presented by N. M. C. Santos at the Tenth Marcel 
Grossmann Meeting on General Relativity}}

\author{T. Barreiro, M. C. Bento, N. M .C. Santos and A. A. Sen}

\address{ Departamento de F\'\i sica, Instituto Superior T\'ecnico,
Av. Rovisco Pais 1, 1049-001 Lisboa, Portugal\\
E-mail:tiago@glencoe.ist.utl.pt, bento@sirius.ist.utl.pt, ncsantos@cfif.ist.utl.pt, anjan@x9.ist.utl.pt}

\maketitle

\abstracts{We use the results from the Wilkinson Microwave Anisotropy Probe
(WMAP) for the locations of  peaks and  troughs of the Cosmic Microwave
Background (CMB) power spectrum, together with  constraints from
large-scale structure, to study a quintessence model in which the pure
exponential potential is modified by a polynomial factor. Our analysis, in
the $(\Omega_m, h, n_s)$ cosmological parameters space 
shows that this quintessence model is
favoured compared to $\Lambda$CDM for $n_s\approx 1$ and relatively high
values of early quintessence; for  $n_s<1$, quintessence and $\Lambda$CDM
give similar results, except for high values of early quintessence, in
which case $\Lambda$CDM is favoured.}

Recent cosmological observations suggest that the dynamics of the Universe at 
the present is dominated by a negative pressure component, called dark energy.
 Various possibilities for the nature of this 
dark energy have been considered, notably the cosmological
 constant and quintessence, a dynamical scalar field
 leading to
 a time-varying equation of state parameter, $w_\phi\equiv p_\phi/\rho_\phi$.
These models most often involve a single field 
\cite{single}
 or, in some cases, two coupled scalar fields \cite{coupled}.
Other possibilities  for the origin of dark energy 
include the generalized Chaplygin gas
 proposal \cite{Chap} and Cardassian models \cite{card}.
In order to unravel the nature of dark energy, it is  crucial
 to use observations so as to be able to discriminate among
 different models. In particular, 
the existence of a dark energy component affects the structure of the CMB power
spectrum, which is particularly sensitive to the amount of dark energy at 
different epochs in cosmology.
 For instance, the
 locations of peaks and troughs 
depend crucially on the amount of dark energy today and at last scattering 
 as well as the dark energy time-averaged equation of state, which are
 model-dependent quantities \cite{Doran1}.  Hence, one can
 use the high-precision measurements recently
 obtained by the BOOMERanG \cite{BOOM}, MAXIMA-1 \cite{MAX}, Archeops 
\cite{ARCH} and, in particular,  WMAP \cite{MAP1} observations to
 constrain dark energy models.

We study the effect of a dark energy component defined
 by the quintessence potential \cite{AS1} 
$V(\phi)=\left[A+\left(\phi-\phi_0\right)^2\right] e^{-\lambda \phi}$
on the location of the first three peaks and the first trough of the CMB power spectrum. 
We have also
 analyzed the consequences of
 cluster abundance constraints \cite{sigma_8const} on  $\sigma_8$, the
 {\it rms} density fluctuations averaged over $8 h^{-1} \textrm{Mpc}$ spheres.
This  M-theory motivated potential leads to  a 
 model with some
 interesting features,
 namely there are 
two types of  attractor solutions giving rise  to an accelerating universe
 today,
 corresponding to permanent or transient
 acceleration \cite{Barrow}. Transient vacuum acceleration is a
 particularly appealing scenario that
 would also solve the apparent incompatibility between an  eternally
accelerating universe 
and  string theory, at least in
its present formulation,  given that  string asymptotic states are
 inconsistent with
spacetimes that exhibit event horizons \cite{Hellerman}.

We should emphasize that restricting the analysis of the CMB power
spectrum to the positions of  peaks and
 troughs, rather than considering the structure of the whole spectrum,
 turns out to be a simple but very powerful tool in constraining the
 model parameters due to the high accuracy  with which these
 positions are now determined, particularly after WMAP results. One
 should also notice that, although our study is limited to the
 $(\Omega_m, h, n_s)$
parameter space, we expect (and have, to some extent, checked) that these are
the most influential parameters; of course, $\Omega_b h^2$ can also be
important
although not within the rather strict WMAP error bars,
 $\Omega_b h^2=0.0224\pm0.0009 $ \cite{MAP1}. 

In our study \cite{bbss}, we used the
accurate analytic approximations given by Doran and Lilley \cite{Doran2} for the positions of the 
first three peaks and first trough. Notice that although those formulae
 were obtained using a standard exponential potential, one expects the results
 to be fairly independent of the form of the potential unless it is
 qualitatively very different from the exponential potential before last scattering.

We found that the dependence of the peaks locations on parameter $A$ is
 extremely small and can be safely neglected. 
As  should be expected, changes in $A$ in order to get 
the transient or permanent acceleration regimes do not alter  the
analysis since the two regimes do not differ significantly until the present,
 and therefore  peak positions should not be affected. For each value of $\lambda$ and $A$, 
$\phi_0$ is chosen such that $\Omega_{tot}=1$.
Hence, the model's behaviour depends essentially on parameter $\lambda$,
 which measures the amount of ``early quintessence'' \cite{Caldwell2003},the  average fraction 
of dark energy before last scattering (${\bar \Omega}_\phi^{ls}\sim 3/\lambda^2$).
 
A lower bound on $\lambda$ already exists from  standard 
Big Bang Nucleosynthesis (BBN)  which implies $\Omega_\phi(\mev)<0.045$,
 or, considering a possible underestimation of systematic errors,
 the more conservative result
 $\Omega_\phi(\mev)<0.09$ 
 \cite{Bean}; these bounds require, respectively,
  $\lambda>9$ and  $\lambda > 6.5$ for 
 the  model we are considering.

We conclude that, with $n_s\approx 1$,
 the $\Lambda$CDM model becomes
increasingly disfavoured compared with this quintessence model 
as the amount of early quintessence becomes higher
 ($\lambda\lsim 15$).
For $n_s < 1$, the opposite is true {\it i.e} $\Lambda$CDM is favoured as
compared to quintessence if
 $\lambda\lsim 15$.
Notice that, as $\lambda$ increases ($\lambda\gsim 18$), independently of the
 value of $n_s$, the
 model's results become
comparable to $\Lambda$CDM's, as should be expected since
 ${\bar \Omega}_\phi^{ls}$ decreases.
Moreover, quintessence is distinguishable from $\Lambda$CDM only
 for $h<0.73$ and $n_s\approx 1$.

Finally, we would like to mention that the
 non-negligible fraction of dark energy at last scattering and during
 structure formation we obtain for this model, typical of early quintessence
 models,  will lead to suppressed clustering
  power on small length scales as suggested by WMAP/CMB/large scale structure
  combined data \cite{Caldwell2003}.

\section*{Acknowledgments}
NMCS acknowledges Funda\c c\~ao para a 
Ci\^encia e a Tecnologia (FCT) under grant SFRH/BD/4797/2001 and Grupo Te\'orico de Altas 
Energias (GTAE) for travel support.

\end{document}